\documentclass[twocolumn,prl,aps,superbib,tightenlines,floatfix]{revtex4}
\usepackage{amsfonts}
\usepackage{amsmath}
\usepackage{amssymb}
\usepackage{graphicx}

\begin{document}

\title{Spin accumulation in ballistic Rashba bar }
\author{J. Yao and Z. Yang\cite{ZY}}
\affiliation{Surface Physics Laboratory (National Key Laboratory), Fudan University,
Shanghai, 200433, China}

\begin{abstract}
We propose an analytic model to study intrinsic spin polarization effect in
a ballistic Rashba bar with two semi-infinite leads. The wave functions
expanded with plane waves in Rashba bar are required to satisfy boundary
conditions at both longitudinal and transverse interfaces. We find
out-of-plane spin Hall accumulation effect can be induced in the Rashba bar
even with large dimensions by injecting \textit{unpolarized} current from
the lead. The longitudinal in-plane spin Hall effect, however, becomes
obscure in large-size sample. An interesting direction-flipping of the
out-of-plane spin accumulation is predicted by altering the Rashba coupling
strength.

\end{abstract}

\maketitle

\input epsf.sty \flushbottom

\textit{Introduction}: An external electric field can be expected to induce
a transverse spin current in non-magnetic semiconductors or result in
out-of-plane spin accumulation near the edges of semiconductor films due to
the spin-orbit coupling (SOC) of electron. It is the so-called spin Hall
effect (SHE)\cite{hirsch}. Very recently, two independent groups have
reported the experimental observation of SHE\cite{kato,wunderlich}, giving
evidence of\textbf{\ }the\textbf{\ }existence of significant SOC in
semiconductors. Such new effect of SOC creates a new way to manipulate
electron spins by means of an electric field, not magnetic field, that may
have great potential applications in future spintronics. SHE was first
proposed by D'yakonov and V. I. Perel' in 1971\cite{yakonov}, then Hirsch%
\cite{hirsch} and Zhang\cite{zhang1}. The existence of SHE effect in p-type
semiconductors and n-type semiconductors in two-dimensional heterostructures
was predicted by Murakami \textit{et al}.\cite{zhang2} and Sinova \textit{%
et al}.\cite{sinova}, respectively. However, the microscopic origin of SHE
still has controversial explanations. The discussion of extrinsic or
intrinsic origin currently attracts many scientists.\cite%
{sheng,hankiewicz,rashba2,loss,shen,inoue2,zhang3,mishchenko,culcer,zhang4}
This letter focuses on the study of spin accumulation in Rashba SOC system
in ballistic region, since ballistic samples usually with small dimensions%
\textbf{\ }are always desirable for future electronic device applications.
References\cite{nikolic2,wang,governale,usaj} investigated spin accumulation
in ballistic system with Rashba interaction. Based on discrete tight-binding
model and Landauer-B\"{u}ttiker formula, Nikoli\'{c} \textit{et al}.\cite%
{nikolic2} numerically demonstrated opposite spin accumulation near the two
edges of the two dimensional semiconductor structure. The phenomena is
qualitatively similar to the experimental observations\cite{kato,wunderlich}%
. Very recently, Wang \textit{et al}.\cite{wang} reported current induced
local spin polarization due to Rashba SOC in a two dimensional narrow strip,
where no lead is considered to attach to the Rashba strip. They concluded
that out-of-plane spin polarization is important mainly in systems with
mesoscopic sizes, and it appears not to be associated with the SHE in bulk
samples.

In this paper, we propose an analytic model to study intrinsic spin
polarization effect in a Rashba bar connected to two semi-infinite leads.
The geometry studied is similar to that proposed by Nikoli\'{c} \textit{et al%
}.\cite{nikolic2}. We, however, don't use the discrete tight-binding model,
but a continuous model to find out its exact solution to Schr\"{o}dinger
equation. The method we developed can properly deal with the coupling
between channels in the Rashba bar. Therefore, the behavior of spin
accumulation as a function of the width of the Rashba bar can be studied
systematically. We find spin accumulation can be induced in Rashba bar by
injecting \textit{unpolarized} current from the lead. With the increase of
the width of the Rashba bar, the peaks of spin accumulation shift towards
the two lateral edges and the oscillations within the bar decrease,
indicating the polarization becomes more prominent in the ballistic bar.
This conclusion is very different from that obtained by Wang \textit{et al}.%
\cite{wang}. Meanwhile, we also report an interesting direction-flipping of
the out-of-plane spin accumulation by altering the spin-orbital coupling
strength, which may be considered as a signature of SHE in ballistic region.

\textit{Theoretical Approach}: We consider a 2DEG Rashba bar with two
opposite interfaces connected to two semi-infinite ideal leads respectively
along $x$ direction. To simplify the calculation (not losing the essential
physics), the leads and the Rashba bar are assumed to have the same widths
in $y$ direction. And open boundary condition is applied at the two lateral
edges of the strip. An electron wave is injected from the right lead to the
left one and crossing the middle Rashba bar. We solve separately Schr\"{o}%
dinger equations of the three regions. The Hamiltonian of the structure is
described by
\begin{equation}
\mathbf{H}=\left[ -\frac{\hbar ^{2}}{2m^{\ast }}\nabla ^{2}+V(y)\right]
\begin{pmatrix}
1 & 0 \\
0 & 1%
\end{pmatrix}%
+\frac{\alpha }{\hbar }(\overrightarrow{\sigma }\times \overrightarrow{p}%
)_{z},  \label{hamil}
\end{equation}%
where $m^{\ast }$ is effective mass of electrons and $\alpha $ the Rashba
SOC strength that can be tuned by gate voltage\cite{nitta}. $\alpha =0$ in
leads, and $\alpha \neq 0$ in the Rashba bar. $V(y)$ is the confined
potential in $y$ direction. For the open boundary, it is simply taken as $%
V(y)=0,$ $(0<y<b);V(y)=\infty ,$ $(y\leq 0;y\geq b)$, where $b$ is the width
of the strip. The total wave function contains two components\textbf{, }$%
\psi (x,y)=\psi _{1}(x,y)\tbinom{1}{0}+\psi _{2}(x,y)\tbinom{0}{1},$ where $%
\psi _{1},\psi _{2}$ can generally expressed as $\psi _{1}(x,y)={\sum }%
_{k_{n},k_{x}}C_{n}(k_{x})\chi _{n}(y)e^{ik_{x}x},\psi _{2}(x,y)={\sum }%
_{k_{n},k_{x}}D_{n}(k_{x})\chi _{n}(y)e^{ik_{x}x}\ $, where $\chi
_{n}(y)=\sin (k_{n}y),$ $n=1,2,3....$ In the formula, $k_{n}=n\pi /b$ and $n$
is the channel number. They are the wave functions that satisfy the
hard-wall boundary conditions in $y$\ direction, $\psi (x,y)|_{y=0;\text{ }%
y=b}=0$. With above wave functions, the Schr\"{o}dinger equation is
simplified as:
\begin{eqnarray}
E_{n}(k_{x})C_{n}(k_{x})+i\alpha \underset{n^{\prime }}{\sum }[\delta
_{nn^{\prime }}k_{x}-\Delta _{n,n^{\prime }}k_{n^{\prime }}]D_{n^{\prime
}}(k_{x}) =0, ~\label{eq1} \\
E_{n}(k_{x})D_{n}(k_{x})-i\alpha \underset{n^{\prime }}{\sum
}[\delta _{nn^{\prime }}k_{x}+\Delta _{n,n^{\prime }}k_{n^{\prime
}}]C_{n^{\prime }}(k_{x}) =0,~ \label{eq2}
\end{eqnarray}%
where $E_{n}(k_{x})$$=$$e_{n}(k_{x})$$-E,~e_{n}(k_{x})$$=$${\hbar ^{2}}%
(k_{x}^{2}$$+$$k_{n}^{2})/{2m}^{\ast }$, $n,n^{\prime }$$=$$1,2,3...,N,$ and
$\Delta _{n,n^{\prime }}$$=$$\frac{2}{b}$${\int_{0}^{b}}$$\sin (k_{n}y)\cos
(k_{n^{\prime }}y)dy$, $(n\neq n^{\prime })$. N is the total number of
channels existing in the Rashba bar and dependent on the width $b$ of the
strip and the energy $E$ of incident electrons. At fixed energy $E$, the
total number $N$ is large with large width $b$. The terms containing $\Delta
_{n,n^{\prime }}$ in Eq.(\ref{eq1}) and (\ref{eq2}) indicate explicitly the
mixing interactions between different subbands due to the Rashba coupling
and the boundaries existing in $y$ direction. Our investigation reveals that
no spin accumulation can be observed for the case of single channel in the
bar, while the accumulation appears in multi-channel cases. This implies the
mixing terms between subbands play an essential role in forming the spin
accumulation effect.

From Eq.(\ref{eq1}) and (\ref{eq2}), we have a polynomial equation
of $k_{x}$
with the highest power of $4N$. Therefore, $4N$ solutions of $k_{x}^{(j)}$ $%
(j=1\rightarrow 4N)$ can be obtained from the equation for any
fixed $E$ and $N$. In the case of single channel, where the SOC
lifts the degenerate spin up and down states at $\pm k_{x}$ into 4
different wave vectors $\pm k_{x1}$ and $\pm k_{x2}$. In general,
$N$ subbands would produce $4N$
different $k_{x}^{(j)}.$ Each $k_{x}^{(j)}$ corresponds to an eigenfunction $%
\widehat{\Phi }_{N}^{(j)}$ which has a length of $2N$. All the coefficients $%
\{C_{n}(k_{x}^{(j)}),D_{n}(k_{x}^{(j)})\}$ in $\widehat{\Phi }_{N}^{(j)}$
are determined up to a normalized constant. And the wave function in Rashba
region with eigenvalue $E$ must be a linear combination of $4N$
eigenfunctions $\widehat{\Phi }_{N}^{(j)},j=1,2,\cdots ,4N$. The combined
coefficients can be denoted by coefficients $\Lambda (k_{x}^{(j)})$.
Therefore, in Rashba region, there are totally $4N$ coefficients $\{\Lambda
(k_{x}^{(j)})\}$ to be determined. If the Rashba bar is at an equilibrium
state, $4N$ eigenfunctions in Rashba bar have equivalent weight. There will
be no spin polarization in the sample since Rashba interaction does not
change the time-reversal symmetry. The time-reverse symmetry, however, is
broken in the bar if there is steady current flowing along $x$ direction.
And spin polarization will occur in the sample. Suppose an \textit{%
unpolarized } incident electron wave injected from the right lead. There are
$2N$ coefficients describing reflected waves in the right lead with respect
to $N$ channels for spin up and down states. Similarly, additional $2N$
coefficients are introduced to express the transmitted waves in the left
lead. Therefore, there are totally $8N$ coefficients to be determined. They
can be solved by boundary conditions at the two interfaces.

The interfaces between leads and Rashba bar denoted as $x=0,a$ are
simplified by a $\delta $ potential\cite{matsuyama} $V_{0}\delta
(x-x_{0}),x_{0}=0,a$ ($a$ is the length of the Rashba bar). The continuous
conditions of wave functions and particle current crossing each interface,
say at $x=0$, yield equations: $\psi _{L}(x)|_{x=0^{-}}=\psi
_{mid}(x)|_{x=0^{+}}$ and$\frac{\widehat{P}_{Lx}}{2m_{L}^{\ast }}\psi
_{L}(x)|_{x=0^{-}}-\frac{\widehat{P}_{mid\text{ }x}}{2m_{mid}^{\ast }}\psi
_{mid}(x)|_{x=0^{+}}=i\frac{V_{0}}{\hbar }\psi _{L}(0),$ where $\widehat{P}%
_{Lx}=(\hbar /i)\partial /{\partial x}$, and $\widehat{P}_{mid\text{ }x}=({%
\hbar }/{i})\partial /{\partial x}-(\alpha m^{\ast }/\hbar )\sigma _{y}$\cite%
{zulicke,matsuyama}. It has been assumed that there's no spin-flip across
the interfaces\cite{matsuyama}. By means of the orthonormality of $\chi _{n}$%
, each subband of spin-up or -down state would satisfy above boundary
conditions. Therefore, for the structure with $N$ channels, two interfaces
along $x$ direction will yield $8N$ equations, the same as the number of
coefficients to be determined. All the coefficients in the wave functions
thus can be exactly solved. With the wave function obtained, the local spin
polarization $\langle S_{i}(x,y)\rangle =\frac{\hbar }{2}\langle \psi
(x,y)|\sigma _{i}|\psi (x,y)\rangle $ is straightforward to be calculated in
the Rashba region.

\textit{Results and Discussion}: In the calculation, we focus on the linear
region and simply assume that the Fermi levels and the effective masses of
electrons in Rashba bar and two leads are the same. They are taken as 10 meV
and 0.04 m$_{e}$, typical values in 2DEG\cite{mireles,nitta}. The current
injection has been normalized to one non-polarized electron wave for each
channel. Meanwhile, we do the statistical average for the spin polarization
of each incident electron wave with energy $E$\ in each channel. The spin
polarizations of $\langle S_{z}\rangle $, $\langle S_{x}\rangle $ and $%
\langle S_{y}\rangle $ with different numbers of channel are demonstrated in
Fig.1, where average integrals along $x$ direction are taken. Note that the
scale in the horizontal axis is normalized in order to compare explicitly
the trends. The channel numbers of 3, 6, and 9 correspond to the widths of
about 120, 210, and 300 nm, respectively. The length of Rashba bar in the
direction of current flow is fixed at 210 nm. It can be found that these
sizes are within the phase-coherent length ($L_{\phi }\lesssim 0.4\mu -1\mu $%
)\cite{nitta,nomura}. Experimentally permitted value of 2.9$\times $10$^{-11}
$ eVm\cite{nitta,engels} is taken for the strength of Rashba coupling. In
the case of 3 channels, $\langle S_{z}\rangle $ has two high peaks with
inverse symmetry distribution along y direction and there are small
oscillations in the middle of the sample. With the increase of the channel
number, the two high peaks become narrow and shift gradually to the lateral
edges with the inside oscillations tending to be smoothed out. The
phenomenon of out-of-plane spin polarization with opposite sign near the two
edges is very much similar to the prediction of spin Hall effect by Hirsch%
\cite{hirsch} and recent experimental observations\cite{kato,wunderlich}.
\begin{figure}[tbp]
\includegraphics*[width=9.0cm]{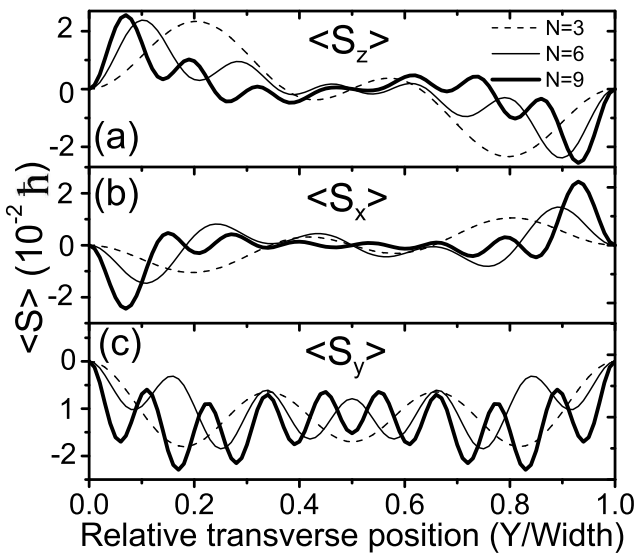}
\caption{The average spin accumulation of $\langle S_{z}\rangle $ (a)$,$ $%
\langle S_{x}\rangle $ (b) and $\langle S_{y}\rangle $ (c) as a function of
the lateral position in the Rashba bar. The spin polarizations varied with
the number of channel are indicated with different curve styles. }
\end{figure}
We also calculated spin polarization of the Rashba bar having 20 channels
(about 630 nm in $y$ direction). The envelop of its accumulation has the
same behavior and keeps very strong spin Hall accumulation effect.
Therefore, the out-of-plane spin accumulation becomes more prominent with
the increase of the channel number in the sample, namely, the width of the
sample. This tendency is different from the conclusion obtained by Wang
\textit{et al}.\cite{wang}. They reported that the out-of-plane polarization
increases at small width of sample, decreases, however, at large width with $%
L\sim 16/k_{F}$ (about 6 channels). From our result, no evidence is found to
support that their such tendency\cite{wang} can also exist in the Rashba bar
with two leads at least to the width of the bar with 20 channels.

The in-plane spin accumulations of $\langle S_{x}\rangle $ and $\langle
S_{y}\rangle $ are given in Fig.1(b) and (c), respectively. The distribution
of $\langle S_{x}\rangle $ as a function of transverse position shares the
same feature of $\langle S_{z}\rangle $, also forming SHE effect\cite%
{hirsch,kato,wunderlich}. With the increase of the channel number, the
effect of $\langle S_{x}\rangle $ also becomes more obvious. For $\langle
S_{z}\rangle $ and $\langle S_{x}\rangle $, the net total spin polarization
is zero because of the inverse symmetry of the distribution along $y$
direction. The spin accumulation of $\langle S_{y}\rangle $, however, has a
uniform sign along the $y$ direction, which gives a net total spin
polarization. The net transverse in-plane magnetization induced from
unpolarized current can be ascribed to the combined factors of Rashba SOC
and time-reversal symmetry breaking in the structures studied.
\begin{figure}[tbp]
\includegraphics[width=8.0cm]{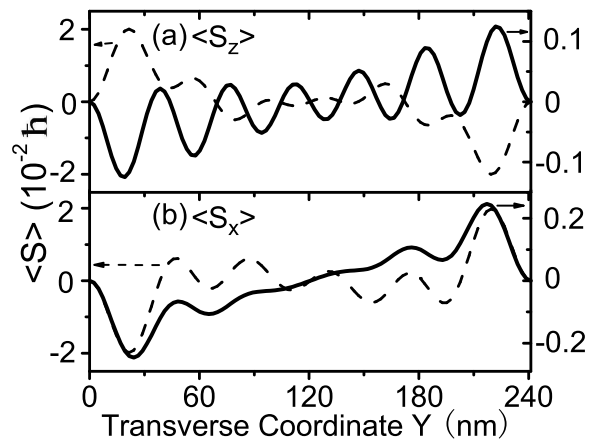}
\caption{Spin accumulation of $\langle S_{z}\rangle $ (a) and $\langle
S_{x}\rangle $ (b) with two different Rashba strengths (2.9$\times $10$%
^{-12} $ eVm, dotted line; 2.9$\times $10$^{-11}$eVm solid line). The
interesting direction-flipping is observed for $\langle S_{z}\rangle $,
while no flipping is obtained for $\langle S_{x}\rangle .$ }
\end{figure}

From Eq. (2) and (3), the behavior of spin polarizations of $\langle
S_{z}\rangle $, $\langle S_{x}\rangle $, and $\langle S_{y}\rangle $ with
opposite Rashba strength ($-\alpha $)\cite{nikolic2} can be obtained
analytically. In Eq.(2) and (3), an opposite Rashba coefficient $-\alpha $
produces an opposite set of $D_{n}(k_{x}^{j})$, while $C_{n}(k_{x}^{j})$ is
not varied$.$ Therefore, $\psi _{1}|_{-\alpha }=\psi _{1}|_{\alpha }$ and $%
\psi _{2}|_{-\alpha }=-\psi _{2}|_{\alpha }$ (see Eq.(2) and (3)), which
directly yield $\langle S_{z}(x,y)\rangle _{-\alpha }=\langle
S_{z}(x,y)\rangle _{\alpha }$ and $\langle S_{x,y}(x,y)\rangle _{-\alpha
}=-\langle S_{x,y}(x,y)\rangle _{\alpha }$. When the direction of Rashba
interaction is reversed, the in-plane spin accumulation flips the
polarization directions, while the out-of-plane one does not. We
demonstrate, however, that in a ballistic SOC strip, the sign of $\langle
S_{z}\rangle $ can be flipped by modulating the SOC strength, as depicted in
Fig.2(a). The thick curve in Fig.2(a) shows the spin polarization of $%
\langle S_{z}\rangle $ with small Rashba strength. With the increase of the
strength to a certain value, opposite peaks near the lateral edges emerge,
and the amplitudes magnify gradually while the inside oscillations decrease.
The flipping-value of $\alpha $ is revealed to be inversely proportional to
the width of the sample. No flipping, however, is observed for $\langle
S_{x}\rangle $ by modulating the value of Rashba strength in the whole
experimental permitted range\cite{nitta,engels}(see Fig. 2(b)). Since the
Rashba interaction strength in a normal 2DEG can be tuned through gate
voltage, we suggest a direct experimental detection of the spin-flipping of $%
\langle S_{z}\rangle $ crossing a critical value of Rashba SOC.

The spin polarizations for $\langle S_{z}(x,y)\rangle $\ and $\langle
S_{x}(x,y)\rangle $\ with long length (about 660 nm) of the Rashba bar are
demonstrated in Fig. 3 (a) and (b), respectively. With 3 channels in the
bar, very obvious period-like variation of the pattern along $x$ is observed
for both $\langle S_{z}\rangle $\ and $\langle S_{x}\rangle $, giving rise
to the spin precession in the sample due to the SOC. The rough period of the
precession is found to be closely related to the characteristic Rashba
length $L_{SO}=\frac{\hbar }{2m^{\ast }\alpha }$. The rough period will be
shorter if there are more channels in the sample or larger SOC. In Fig.
3(a), the sign of $\langle S_{z}\rangle $\ keeps in period-like variation
along $x$ direction, while $\langle S_{x}\rangle $\ flips its sign during
the variation (see Fig. 3(b)). This featured pattern of $\langle
S_{x}\rangle $, however, is obscured with the increase of the width of the
sample due to having more oscillations along $x$ direction. Therefore, for
large dimensions of ballistic bar, the out-of-plane SHE can survive, while
the longitudinal in-plane spin Hall polarization may not be detectable in
the case.
\begin{figure}[tbp]
\includegraphics*[width=8.8cm]{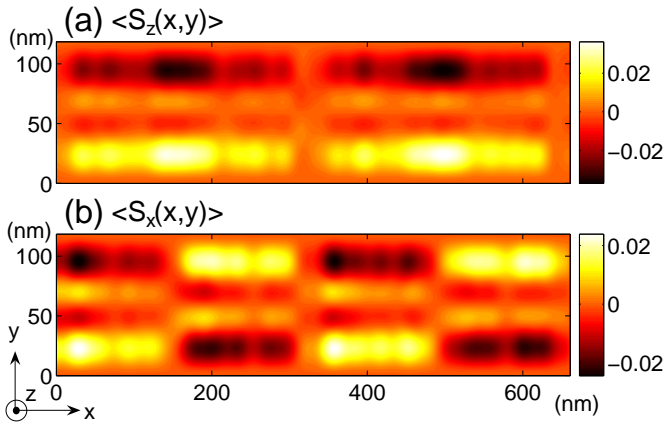}
\caption{Contour plots of spin accumulation of out-of-plane $\langle
S_{z}(x,y)\rangle $ (a) and longitudinal in-plane $\langle S_{x}(x,y)\rangle
$ (b) with long length of rashba bar. There are 3 channels in the sample.
The Rashba strength used is 2.9 $\times $10$^{-11}$ eVm.}
\end{figure}

The variation of spin-polarization pattern under opposite bias voltage\cite%
{kato} is predicted. If the center of the Rashba bar is taken as the origin,
the wave function should have following symmetry: {\small {\ $\langle \psi
(x,y)|S_{i}|\psi (x,y)\rangle _{-V}=\langle \psi (x,y)|P^{\dag }U^{\dag
}S_{i}UP|\psi (x,y)\rangle _{V,}$ } }where $P$ is the rotation of $\pi $
angle around $z$ axis in orbital space, and $U$ is SU(2) rotation with same
angle $\pi $ in spin space. It is straightforward to know that $U=%
\begin{pmatrix}
i & 0 \\
0 & -i%
\end{pmatrix}%
$. Therefore, following relations can be obtained for the case of reverse
bias. {\small
\begin{eqnarray}
\langle \psi (x,y)|S_{z}|\psi (x,y)\rangle _{-V} &=&\langle \psi
(-x,-y)|S_{z}|\psi (-x,-y)\rangle _{V,}  \notag \\
\langle \psi (x,y)|S_{x}|\psi (x,y)\rangle _{-V} &=&-\langle \psi
(-x,-y)|S_{x}|\psi (-x,-y)\rangle _{V,}  \notag \\
\langle \psi (x,y)|S_{y}|\psi (x,y)\rangle _{-V} &=&-\langle \psi
(-x,-y)|S_{y}|\psi (-x,-y)\rangle _{V.}  \notag
\end{eqnarray}%
}Our numerical calculations have exactly confirmed these symmetric
relations, which are not completely same as the ones reported by Nikoli\'{c}
\textit{et al}.\cite{nikolic2}. For $\langle S_{z}(x,y)\rangle $ under
reverse bias voltage, the pattern of spin polarization does not just flip
its sign\cite{kato}, but has inverse symmetry compared to the original one
(see Fig. 3(a)). It is also an intrinsic feature of SHE in ballistic region.

\textit{Summary}: We studied intrinsic spin polarization effect in a
ballistic Rashba bar by proposing an analytic model. Both the out-of-plane
and longitudinal in-plane spin Hall accumulation effects are observed. In
the sample with large dimensions, only the out-of-plane spin Hall
accumulation can be observed obviously. The transverse in-plane spin
accumulation shows same sign across the sample, giving net spin
magnetization. This spin non-conservation can be ascribed to the Rashba
interaction together with the time-reversal symmetry breaking in the
structure. An interesting direction-flipping of the out-of-plane spin
accumulation is found by altering the Rashba coupling strength. This
phenomenon is suggested to be confirmed by experiment. Some symmetric
relations of spin polarizations are predicted.

\textit{Acknowledgments}:\textbf{\ }The authors are grateful to Shou-Cheng
Zhang, Fuchun Zhang and Ruibao Tao for very helpful discussion. The work is
supported by National Natural Science Foundation with grant No. 10304002 and
the Fudan High-end Computing Center.

\end{document}